\documentclass[12pt]{article}

\usepackage{amssymb}
\usepackage{latexsym}
\usepackage{epsfig}
\usepackage{citesort}

\def\ds{\displaystyle}
\def\beq{\begin{equation}}
\def\eeq{\end{equation}}
\def\bea{\begin{eqnarray}}
\def\eea{\end{eqnarray}}
\def\beeq{\begin{eqnarray}}
\def\eeeq{\end{eqnarray}}

\def\vel{\left|}
\def\ver{\right|}
\def\nnb{\nonumber}
\def\ga{\left(}
\def\dr{\right)}

\def\lla{\left<}
\def\rra{\right>}
\def\rar{\rightarrow}
\def\nnb{\nonumber}

\def\ba{\begin{array}}
\def\ea{\end{array}}

\def\xis0{{\Xi^{*0}}}

\def\g5{\gamma_5}

\def\simlt{\stackrel{<}{{}_\sim}}

\begin{document}

%%%%%%%%%%%%%%%%%%%%%%%%%%%
%  Titlepage

\begin{titlepage}

\renewcommand{\thefootnote}{\fnsymbol{footnote}}

\vspace*{-0.5truecm}
\begin{flushright}
{\tt hep-ph/0506188 \\
YITP-05-33}
\end{flushright}
\vspace*{0.5truecm}

\begin{center}

{\Large \bf \boldmath CP violation in the $B \to K \ell^+ \ell^-$ decay}\\

\hspace{10pt}\\
\vspace{1truecm}

{\bf T. M. Aliev$^a$ \footnote{taliev@metu.edu.tr},
S. Rai Choudhury$^b$\footnote{src@physics.du.ac.in},
A. S. Cornell$^c$\footnote{alanc@yukawa.kyoto-u.ac.jp}
and
Naveen Gaur$^b$\footnote{naveen@physics.du.ac.in}}
\vskip .8cm
$^a ${\it Department of Physics, Middle East Technical University ,
06531, Ankara, Turkey}, \\
{\it Institute of Physics, Baku, Azerbaijan.} \\
$^b$ {\it Department of Physics $\&$ Astrophysics,
University of Delhi, Delhi - 110 007, India.} \\
$^c${\it Yukawa Institute for Theoretical Physics,
Kyoto University, Kyoto 606-8502, Japan.}

\vskip 2cm
{\bf Abstract}
\vskip .2cm
\begin{quote}
Standard Model (SM) CP asymmetries in $B \to K \ell^+ \ell^-$ are
expected to be very small. This feature could help in the
understanding of new physics scenarios which predict the existence
of CP odd phases in various Wilson coefficients. In this paper we
have analyzed the $B \to K \ell^+ \ell^-$ decay in beyond the SM
scenarios where the Wilson coefficients have new CP odd phases.
The sensitivity of the CP asymmetries on these new weak phases is
discussed.
\end{quote}
\end{center}

%\vspace{1cm}
~~~PACS numbers: 13.20.He, 12.60.--i, 13.88.+e

\end{titlepage}

%%%%%%%%%%%%%%%%%%%%%%%%%%%%%%%%%%%%%%%%%%%%%%%
%  Section 1: Introduction

\section{Introduction}

One of the key ingredients in the Standard Model (SM) is CP violation,
which can be described by the Cabibbo-Kobayashi-Maskawa (CKM)
matrix\cite{R1}. However, even with this description we still have
an incomplete picture concerning the origin of CP violation in the
SM. The exploitation of CP violation from the theoretical and
experimental sides of physics is very exciting, as it may open a
window to the existence of new physics beyond the SM. Note that the
existence of CP violation is a well established fact in $K$\cite{R3}
and $B$\cite{R4} meson systems.

\par In order to study the sources of CP violation it is promising to
consider those observables which are sensitive to the possible CP
phases. For example, CP asymmetries in decay widths and lepton
polarization asymmetries, such as explored in references
\cite{R5,R7,R8,R10}.

\par One of the promising directions for measuring CP violation is the
analysis of rare semi-leptonic decays. From the experimental
perspective the exclusive decay modes, such as $B \to K \ell^+ \ell^-$
and $B \to K^* \ell^+ \ell^-$, are easy to measure. Two years ago the
Belle\cite{R11} and BaBar\cite{R12} collaborations announced the
following results for the branching ratios for the $B \to K \ell^+
\ell^-$ and $B \to K^* \ell^+ \ell^-$ decays;
\bea
Br(B \rar K \ell^+ \ell^-) = \left\{ \begin{array}{lc}
\left( 4.8^{+1.0}_{-0.9} \pm 0.3 \pm 0.1\right) \times
10^{-7}& \cite{R11}~,\\ \\
\left( 0.65^{+0.14}_{-0.13} \pm 0.04\right) \times 10^{-6}&
\cite{R12}~,\end{array} \right. \nnb
\eea
\bea
Br(B \rar K^* \ell^+ \ell^-) = \left\{ \begin{array}{lc}
\left( 11.5^{+2.6}_{-3.4} \pm 0.8 \pm 0.2\right) \times
10^{-7}& \cite{R11}~,\\ \\
\left( 0.88^{+0.23}_{-0.29}\right) \times 10^{-6}&
\cite{R12}~.\end{array} \right. \nnb
\eea

\par The analysis for study of possible CP violation in $B \to K^*
\ell^+ \ell^-$ was done in earlier works \cite{R13,Kruger:2000zg}.
The goal of our present work is to similarly study the possible CP
violation asymmetry in the exclusive $B \to K \ell^+ \ell^-$ decay
using the most general form of the effective Hamiltonian, including
all possible forms of interactions. Such an analysis will be useful
for comparisons with experimental results, as the inclusive modes are
generally hard to measure. Note that the CP violation in the decay $B
\to K \ell^+ \ell^-$ is induced by the $b \to s \ell^+ \ell^-$
transition, which in the SM is practically equal to zero. This is due to
the CKM factors $V_{ub} V_{us}^*$ being negligible, with the result
that the unitarity condition produces only an overall phase factor in
the matrix element. Therefore the CP asymmetry is strongly
suppressed. As such, any deviation from zero for the CP asymmetry
would be an indication of new physics.

\par This paper is organized as follows. In section 2, using
the most general form of the effective Hamiltonian, we derive the
matrix element of the $B \to K \ell^+ \ell^-$ decay in terms of the $B
\to K$ transition form-factors. We also derive in this section the
general analytic expression for the CP violating asymmetry. Section 3
contains our numerical analysis of the CP violating asymmetries
together with our conclusions.

%%%%%%%%%%%%%%%%%%%%%%%%%%%%%%%%%%%%%%%%%%%%%%%
%  Section 2

\section{The matrix element for the $B \to K \ell^+ \ell^-$ decay}

In this section we calculate the matrix element for the $B \to K
\ell^+ \ell^-$ decay, which is governed by the $b \to s \ell^+
\ell^-$ transition at the quark level. The most general form of
the effective Hamiltonian,  for the $b \to s \ell^+ \ell^-$
transition (in terms of the twelve model independent four-Fermi
interactions) can be written in the following form\cite{R14};
\begin{eqnarray}\label{eff}
{\cal H}_{eff} & = & \displaystyle \frac{\alpha G_F}{\sqrt{2} \pi}
V_{tb} V_{ts}^* \Bigg[ C_{SL} \left( \bar{s} i \sigma_{\mu \nu}
\frac{q^{\nu}}{q^2} L b \right) \bar{\ell} \gamma^{\mu} \ell +
C_{BR} \left( \bar{s} i \sigma_{\mu \nu} \frac{q^{\nu}}{q^2} R b
\right)
\bar{\ell} \gamma^{\mu} \ell \nonumber \\
&& \hspace{0.8in} + C_{LL}^{tot}\left( \bar{s}_L \gamma_{\mu} b_L
\right) \bar{\ell}_L \gamma^{\mu} \ell_L + C_{LR}^{tot} \left(
\bar{s}_L \gamma_{\mu} b_L \right) \bar{\ell}_R \gamma^{\mu} \ell_R
+ C_{RL} \left( \bar{s}_R \gamma_{\mu} b_R \right) \bar{\ell}_L
\gamma^{\mu}
\ell_L \nonumber \\
&& \hspace{0.8in} + C_{RR} \left( \bar{s}_R \gamma_{\mu} b_R \right)
\bar{\ell}_R \gamma^{\mu} \ell_R + C_{LRLR} \left( \bar{s}_L b_R
\right) \bar{\ell}_L \ell_R + C_{RLLR} \left( \bar{s}_R b_L \right)
\bar{\ell}_L \ell_R \nonumber \\
&& \hspace{0.8in} + C_{LRRL} \left( \bar{s}_L b_R \right)
\bar{\ell}_R \ell_L + C_{RLRL} \left( \bar{s}_R b_L \right)
\bar{\ell}_R \ell_L + C_T \bar{s} \sigma_{\mu \nu} b \bar{\ell}
\sigma^{\mu \nu} \ell
\nonumber \\
&& \hspace{0.8in} + i C_{TE} \epsilon^{\mu \nu \alpha \beta} \bar{s}
\sigma_{\mu \nu} b \bar{\ell} \sigma_{\alpha \beta} \ell \Bigg] ,
\label{Hamiltonian}
\end{eqnarray}
where $L/R = \frac{1}{2} ( 1 \mp \gamma_5 )$, the $C_X$'s are the
Wilson coefficients of the four-Fermi interactions and
$q_{\mu}=(p_B-p_K)_{\mu}= (p_+ + p_-)_{\mu}$ is the momentum
transfer. Among the twelve Wilson coefficients several already
exist in the SM. For example, the first two terms with
coefficients $C_{SL}$ and $C_{BR}$ describe the penguin operators,
where in the SM these coefficients are equal to  $-2 m_s
C_7^{eff}$ and $-2 m_b C_7^{eff}$. The next four terms in
Eq.(\ref{eff}) are the vector type interactions with coefficients
$C_{LL}^{tot}$, $C_{LR}^{tot}$, $C_{RL}$ and $C_{RR}$. Two of
these vector interactions, $C_{LL}^{tot}$ and $C_{LR}^{tot}$, also
exist in the SM with the form $(C_9^{eff}-C_{10})$ and
$(C_9^{eff}+C_{10})$. Therefore we can say that the coefficients
$C_{LL}^{tot}$ and $C_{LR}^{tot}$ describe the sum of the
contributions from the SM and the new physics, where they can be
written as; \bea
C_{LL}^{tot} &=& C_9^{eff} - C_{10} + C_{LL}~, \nnb \\
C_{LR}^{tot} &=& C_9^{eff} + C_{10} + C_{LR}~. \nnb
\eea
The terms with coefficients $C_{LRLR}$, $C_{RLLR}$, $C_{LRRL}$ and
$C_{RLRL}$ describe the scalar type interactions. The last two terms,
with the coefficients $C_T$ and $C_{TE}$, describe the tensor type
interactions.

\par Now that we have the effective Hamiltonian, describing the $b \to
s \ell^+ \ell^-$ decay at a scale $\mu\simeq m_B$, we can write down
the matrix elements for the $B \to K \ell^+ \ell^-$ decay. The matrix
element for this decay can be obtained by 
sandwiching the effective Hamiltonian between $B$ and $K$ meson
states; which are parameterized in terms of form-factors which depend
on the momentum transfer squared, $q^2=(p_B-p_K)^2= (p_+-p_-)^2$. It
follows from Eq.(\ref{eff}) that in order to calculate the amplitude
of the $B \to 
K \ell^+ \ell^-$ decay the following matrix elements are required;
$$\lla K\vel \bar s \gamma_\mu b \ver B \rra ,
\lla K \vel \bar s i\sigma_{\mu\nu} q^\nu b \ver B \rra ,
\lla K \vel \bar s b \ver B \rra ,
\lla K \vel \bar s \sigma_{\mu\nu} b \ver B \rra . $$
These matrix elements are defined as follows \cite{R16,R17,R20};
\bea \label{e6002}
\lla K(p_{K}) \vel \bar s \gamma_\mu b \ver B(p_B) \rra = f_+
\Bigg[ (p_B+p_K)_\mu - \frac{m_B^2-m_K^2}{q^2} \, q_\mu \Bigg] +
f_0 \,\frac{m_B^2-m_K^2}{q^2} \, q_\mu , \eea

 \bea \label{e6003} \lla K(p_{K}) \vel \bar s
\sigma_{\mu\nu}
 b \ver B(p_B) \rra = -i \, \frac{f_T}{m_B+m_K}
\Big[ (p_B+p_K)_\mu q_\nu - q_\mu (p_B+p_K)_\nu\Big]~. \eea
 Note
that the finiteness of Eq.(\ref{eff}) at $q^2=0$ is guaranteed by
assuming that $f_+(0) = f_0(0)$.
\par The matrix elements $\lla K(p_{K}) \vel \bar s i \sigma_{\mu\nu}
q^\nu b \ver B(p_B) \rra$ and $\lla K \vel \bar s b \ver B \rra$ can
be obtained from Eqs.(\ref{e6002}) and (\ref{e6003}) by multiplying
both sides of these equations by $q^\mu$ and using the equations of
motion, we get;
\bea \label{e6004} \lla
K(p_{K}) \vel \bar s b \ver B(p_B) \rra & = &
f_0 \, \frac{m_B^2-m_K^2}{m_b-m_s} , \\
\label{e6005} \lla K(p_{K}) \vel \bar s i \sigma_{\mu\nu} q^\nu b
\ver B(p_B) \rra & = & \frac{f_T}{m_B+m_K} \Big[ (p_B+p_K)_\mu q^2 -
q_\mu (m_B^2-m_K^2) \Big] . \eea

\par As we have already mentioned the form-factors entering
Eqs.(\ref{e6002})-(\ref{e6005}) represent the hadronization
process, where in order to calculate these form-factors
information about the nonperturbative region of QCD is required.
Therefore for the estimation of the form-factors to be reliable a
nonperturbative approach is needed. Among the nonperturbative
approaches the QCD sum rule \cite{R15} is more predictive in
studying the properties of hadrons. The form-factors appearing in
the $B \to K$ transition are computed in the framework of the
three point QCD sum rules \cite{R16} and  in the light cone QCD
sum rules\cite{R17,R20}. We will use the result of the work
in \cite{R20} where radiative corrections to the leading twist
wave functions and $SU(3)$ breaking effects are taken into
account. As a result the form-factors are parameterized in the
following way \cite{R20};  
\beq\label{fi}
f_i(q^2)=\frac{r_1}{1-q^2/m_1^2}+\frac{r_2}{(1-q^2/m_1^2)^2} ,
\eeq where $1=+$ or $T$, and \beq\label{f0}
f_0(q^2)=\frac{r_2}{1-q^2/m_{fit}^2} , \eeq with $m_1=5.41$GeV
and the other parameters as given in Table 1.
%%%%%%%%%%%%%%%%%%%%%%%%%%%%%%%%%
%  Table
\begin{table}[htb]
\begin{center}
\begin{tabular}{|c||c|c|c|}
  \hline
   & $r_1$ &$r_2$ &$m_{fit}^2$ \\
  \hline \hline
 $f_+ $&$ 0.162$ &$0.173$ &$--$  \\
  $f_0$ & 0. &$ 0.33$ &37.46 \\
  $f_T$&$ 0.161 $&$0.198$& $--$  \\
  \hline
\end{tabular}
\caption{\it The parameters for the form-factors of the $B \to K$
transition as given in \cite{R20}.}\label{app:table:1}
\end{center}
\end{table}
%%%%%%%%%%%%%%%%%%%%%%%%%%%%%%%%%

\par Using the definition of the form factors given in
Eqs.(\ref{e6002})-(\ref{e6005}) we arrive at the following matrix
element for the $B \to K \ell^+ \ell^-$ decay;
\bea \label{e6006} {\cal M}(B\rightarrow K
\ell^{+}\ell^{-}) & = & \frac{G_F \alpha}{4 \sqrt{2} \pi} V_{tb}
V_{ts}^\ast \Bigg\{ \bar \ell \gamma^\mu \ell \, \Big[
A (p_B+p_K)_\mu + B q_\mu \Big] \nnb \\
&&\hspace{1in}+ \bar \ell \gamma^\mu \gamma_5 \ell \, \Big[
C (p_B+p_K)_\mu  + D q_\mu \Big]
+\bar \ell \ell \,Q
+ \bar \ell \gamma_5 \ell \, N \nnb \\
&&\hspace{1in}+ 4 \bar \ell \sigma^{\mu\nu}  \ell\, (- i G )
\Big[ (p_B+p_K)_\mu q_\nu - (p_B+p_K)_\nu q_\mu
\Big] \nnb \\
&&\hspace{1in}+ 4 \bar \ell \sigma^{\alpha\beta}  \ell \,
\epsilon_{\mu\nu\alpha\beta} \, H
\Big[ (p_B+p_K)_\mu q_\nu - (p_B+p_K)_\nu q_\mu \Big]
\Bigg\} .
\eea

\par The functions entering Eq.(\ref{e6006}) are defined as;
\bea
\label{e6007} A &=& (C_{LL}^{tot} + C_{LR}^{tot} + C_{RL} +
C_{RR}) f_+ +
2 (C_{BR}+C_{SL}) \frac{f_T}{m_B+m_{K}} , \nnb \\
B &=& (C_{LL}^{tot} + C_{LR}^{tot}+ C_{RL} + C_{RR}) f_- -
2 (C_{BR}+C_{SL})\frac{f_T}{(m_B+m_{K})q^2}(m_B^2-m_K^2) , \nnb \\
C &=& (C_{LR}^{tot} + C_{RR} - C_{LL}^{tot} - C_{RL}) f_+ ,\nnb \\
D &=& (C_{LR}^{tot} + C_{RR} - C_{LL}^{tot} - C_{RL}) f_- , \nnb \\
Q &=& f_0 \frac{m_B^2-m_K^2}{m_b-m_s}(C_{LRLR} + C_{RLLR}+C_{LRRL} +
C_{RLRL}) ,\nnb \\
N &=& f_0 \frac{m_B^2-m_K^2}{m_b-m_s}(C_{LRLR} + C_{RLLR}-C_{LRRL} -
C_{RLRL}) ,\nnb \\
G &=& \frac{C_T}{m_B+m_K} f_T ,\nnb \\
H &=& \frac{C_{TE}}{m_B+m_K} f_T ,
\eea
where
\bea
f_- = (f_0-f_+) \frac{m_B^2-m_K^2}{q^2}~.\nnb
\eea

\par From Eq.(\ref{e6006}) it follows that the difference from the SM
is due to the last four terms only, namely the scalar and tensor type
interactions. For an analysis of the CP asymmetry it is necessary to
compute the differential decay width for $B \to K \ell^+ \ell^-$. From
the expression of the matrix element given in Eq.(\ref{e6006}) we
calculate the following result for the dilepton invariant mass
spectrum;
\bea \label{e6008} \frac{d\Gamma}{d\hat{s}}(B \rar K \ell^+ \ell^-)
= \frac{G^2 \alpha^2 m_B}{2^{14} \pi^5} \vel V_{tb}V_{ts}^\ast
\ver^2 \lambda^{1/2}(1,\hat{r}_K,\hat{s}) v \Delta(\hat{s})~, \eea
where
$\lambda(1,\hat{r}_K,\hat{s})=1+\hat{r}_K^2+\hat{s}^2-2\hat{r}_K-2\hat{s}-
2\hat{r}_K\hat{s}$, $\hat{s}=q^2/m_B^2$, $\hat{r}_K=m_K^2/m_B^2$,
$\hat{m}_\ell=m_\ell/m_B$, $v=\sqrt{1-4\hat{m}_\ell^2/\hat{s}}$ is the
final lepton velocity, and $\Delta(\hat{s})$ is;
\bea
\label{e6009} \Delta & = & \frac{4 m_B^2}{3} \mbox{\rm
Re}\Big[ -96 \lambda m_B^3 \hat{m}_\ell (A G^\ast) + 24 m_B^2
\hat{m}_\ell^2 (1-\hat{r}_K) (C D^\ast) +
12 m_B \hat{m}_\ell (1-\hat{r}_K) (C N^\ast) \nnb \\
&& + 12 m_B^2 \hat{m}_\ell^2 \hat{s} \,\vel D \ver^2 +
3 \hat{s} \,\vel N \ver^2 + 12 m_B \hat{m}_\ell \hat{s} (D N^\ast)
+ 256 \lambda m_B^4 \hat{s} v^2  \,\vel H \ver^2
+ \lambda m_B^2 (3-v^2)\,\vel A \ver^2 \nnb \\
&& +s 3 \hat{s} v^2  \,\vel Q \ver^2
+ 64 \lambda m_B^4 \hat{s} (3-2 v^2)\,\vel G \ver^2
+ m_B^2 \big\{2 \lambda - (1-v^2)\big[ 2 \lambda -
3(1-\hat{r}_K)^2 \big] \big\}\,\vel C \ver^2 \Big]~.
\eea

\par As we have already mentioned, our goal in this work is the study 
of possible CP violating asymmetries beyond the SM in the $B \to K
\ell^+ \ell^-$ decay; at this point we shall briefly remind the reader
of the situation in the SM. In the SM the $C_9$ Wilson coefficient is
the only one to have 
strong and weak phases. Strong phases arise from the short distance
effects and resonances whereas the weak phase comes from the CKM 
elements. The remaining two coefficients, $C_7$ and $C_{10}$, are
strictly real within the SM. From the parameterization of the
form-factors it follows 
that they are inherently real and thus the imaginary parts in the
functions in Eq.(\ref{e6009}) can come only from the Wilson
coefficients in Eq.(\ref{eff}). By {\it strong} and {\it weak} phases
we mean the phases which are CP even and odd respectively.
In other words we shall consider the picture where CP
violating effects due to the short distance dynamics are parameterized
by the Wilson coefficients. In principle all Wilson coefficients can
have nonzero strong and weak phases . 
%%%%%%%%%%%
\par In general the amplitude for $\bar{B} \to K$ has the general form
\cite{Kruger:2000zg};
\beq
A(\bar{B} \to K) = e^{i \phi_1} A_1 e^{i \delta_1}
+ e^{i \phi_2} A_2 e^{i \delta_2} ,
\label{cp1}
\eeq
where the strong phases are labeled as $\delta$'s and the weak phases
by $\phi$'s. As noted above the strong phases are CP even, whereas
weak phases are odd under CP. Thus we arrive at an amplitude for the
conjugated process, $B \to \bar{K}$, from Eq.(\ref{cp1}); 
\beq
\bar{A}(B \to \bar{K}) = e^{-i \phi_1} A_1 e^{i \delta_1}
+ e^{-i \phi_2} A_2 e^{i \delta_2} , 
\label{cp2}
\eeq
where the amplitudes of the decay rate of particle and anti-particle
can be defined by the CP asymmetry (in the decay rate) as; 
\beq
A_{CP} = \frac{ |A|^2 - |\bar{A}|^2}{ |A|^2 + |\bar{A}|^2}
=  
\frac{- 2 A_1 A_2 Sin(\phi_1 - \phi_2) Sin(\delta_1 - \delta_2)}{A_1^2
+ 2 A_1 
A_2 Cos(\phi_1 - \phi_2) Cos(\delta_1 - \delta_2) + A_2^2} .
\label{cp3}
\eeq
Note that from the above expression we observe that in order to have
CP asymmetry we should have both strong and weak phases in the
amplitude; where the strong phases are provided by $C_9^{eff}$. In the SM the weak phases for
the $b \to s \ell^+ \ell^-$ transition are negligible and hence the CP
asymmetry for processes based on the quark level transitions, $b \to s
\ell^+ \ell^-$, are highly suppressed. We will now consider the CP
asymmetry in the decay width which is defined as;
\beq\label{acp}
A_{CP}(q^2)= \frac{\displaystyle
\frac{d\Gamma}{d\hat{s}}(\overline{B} \to K \ell^+ \ell^-)
-\frac{d\Gamma}{d\hat{s}}(B \to
\overline{K} \ell^+ \ell^-)}{\displaystyle
\frac{d\Gamma}{d\hat{s}}(\overline{B} \to K \ell^+ \ell^-)
+ \frac{d\Gamma}{d\hat{s}}(B \to \overline{K} \ell^+ \ell^-)} .
\eeq
Note that one can also have a CP asymmetry from the Forward-Backward
(FB) asymmetry\cite{Choudhury:1997xa}. However, in our present case
the FB asymmetry for $B \to K \ell^+ \ell^-$ vanishes within the SM.  

\par We shall now consider the minimal extension of these Wilson
coefficients. In this approach we shall assume that the Wilson
coefficients corresponding to scalar and tensor type interactions
vanish identically (of course in the general case we can consider all
Wilson coefficients with an arbitrary weak phase). For scalar type
operators which emerge in Supersymmetric (SUSY) models and two Higgs
Doublet Models (2HDM) this assumption is justified when we have
electrons or muons in the final state. The reason being that in SUSY
and 2HDM these operators originate from an Higgs exchange which
results in Wilson coefficients which are proportional to $m_\ell$, and
hence negligible for $\ell = e, \mu$.

\par The Wilson coefficients for the dipole operator obeys;
\beq C_{BR}=-2 C_7^{eff} m_b \quad , \quad
C_{SL} = - 2 C_7^{eff} m_s ,
\eeq
with
$$ C_7^{eff} = |C_7^{eff}| exp(i \phi_7) , $$
where $\phi_7$ is an arbitrary phase and it is not constrained by the
already observed branching ratio $Br(B\to K^* \gamma)$.

\par Regarding the appearance of the new weak phase in $C_{10}$ we
feel that a few words are in order. One of the possible discrepancies
between the experimental results\cite{R21} and the theoretical
prediction for $B\to \pi K$ (from the $B \to \pi \pi$ data) can be
resolved, 
as proposed in \cite{R22}, by introducing a complex phase in the
Wilson coefficient $C_{10}=C_{10}^{SM} \exp(i\phi_{10})$.
%, where $\phi_{10}=\frac{103}{180} \pi$. 
In this prescription the weak phase
given to $C_{10}$ does not effect the CP asymmetry in $B \to K \ell^+
\ell^-$.

\par We will assume that the Wilson coefficients $C_{RL}$ and $C_{RR}$
also have weak phases, that is;
\bea
C_{RL}&=& |C_{RL}| \exp(i\phi_{RL}) , \nonumber\\
C_{RR}&=& |C_{RR}| \exp(i\phi_{RR}) .
\eea

\par The Wilson coefficient $C_{9}^{eff}(m_{b}, q^{2})$ has a finite
phase, where, in order to better appreciate this, we write its
explicit phase content as;
\begin{eqnarray}
\label{c9} C_{9}^{eff}(m_{b})=C_{9}(m_{b})\Bigg\{ 1 + \frac{\alpha_s
\ga \mu \dr }{\pi} \omega \ga \hat s \dr \Bigg\} + Y_{SD}\ga m_{b},
\hat s \dr+Y_{LD} \ga m_{b}, \hat s \dr~ ,
\end{eqnarray}
where $C_{9}(m_{b})=4.334$. Here $\omega \ga \hat s \dr$ represents
the ${\cal{O}}(\alpha_{s})$ corrections coming from the four quark
operator ${\cal{O}}_{9}$ \cite{R23};
\bea \omega \ga
\hat s \dr &=& - \frac{2}{9} \pi^2 - \frac{4}{3} Li_2 \ga \hat s
\dr - \frac{2}{3} \ln \ga \hat s\dr \,\ln \ga 1 -\hat s \dr
- \,\frac{5 + 4 \hat s}{3 \ga 1 + 2 \hat s \dr} \ln \ga 1 -\hat s \dr
\nnb \\
&&-\frac{2 \hat s \ga 1 + \hat s \dr \ga 1 - 2 \hat s \dr} {3 \ga 1
- \hat s \dr^2 \ga 1 + 2 \hat s \dr} \, \ln \ga \hat s\dr + \frac{5
+ 9 \hat s - 6 {\hat s}^2} {3 \ga 1 - \hat s \dr \ga 1 + 2 \hat s
\dr}~. \eea

\par In Eq.(\ref{c9}) $Y_{SD}$ and $Y_{LD}$ represent, respectively,
the short and long distance contributions to the four quark operators
${\cal{O}}_{i=1,\cdots,6}$ \cite{R23,R24}. Here $Y_{SD}$ can be
obtained by a perturbative calculation;
\bea Y_{SD}\ga
m_{b}, \hat s \dr &=& g \ga \hat m_c,\hat s \dr
\left[3 C_1 + C_2 + 3 C_3 + C_4 + 3 C_5 + C_6 \right] \nnb \\
&&- \frac{1}{2} g \ga 1,\hat s \dr
\left[4 C_3 +4 C_4 + 3 C_5 + C_6 \right] \nnb \\
&&- \frac{1}{2} g \ga 0,\hat s \dr \left[ C_3 + 3  C_4 \right]
+ \frac{2}{9} \left[ 3 C_3 + C_4 + 3 C_5 + C_6 \right] \nnb \\
&&- \frac{V_{us}^* V_{ub}}{V_{ts}^* V_{tb}} \left[ 3 C_1 + C_2
\right] \left[ g \ga 0,\hat s \dr - g \ga \hat m_c,\hat s \dr
\right]~, \eea
where the loop function $g \ga m_q, s \dr$ represents the loops of
quarks with mass $m_{q}$ at the dilepton invariant mass $s$. This
function develops absorptive parts for dilepton energies $s= 4
m_q^{2}$;
\bea g \ga \hat m_q,\hat s \dr & = & -
\frac{8}{9} {\rm ln} \hat m_q + \frac{8}{27} + \frac{4}{9} y_q -
\frac{2}{9} \ga 2 + y_q \dr \sqrt{\vel 1 - y_q \ver} \nnb \\
&&\times \Big\{ \Theta(1 - y_q) \ga {\rm ln} \frac{1  + \sqrt{1 -
y_q}}{1  -  \sqrt{1 - y_q}} - i \pi \dr + \Theta(y_q - 1) 2 \,{\rm
arctan} \frac{1}{\sqrt{y_q - 1}} \Big\}, \eea
where  $\hat m_q= m_{q}/m_{b}$ and $y_q=4 \hat m_q^2/\hat
s$. Therefore, due to the extension of the absorptive parts of $g \ga
\hat m_q,\hat s \dr$ we see that the strong phases come from
$Y_{SD}$. In particular one notices that the terms proportional to $g
\ga 0,\hat s \dr$ have a non-vanishing imaginary part independent of
the dilepton invariant mass.

\par In addition to these perturbative contributions the $\bar{c}c$
loops can excite low-lying charmonium states $\psi(1s), \cdots,
\psi(6s)$ whose contributions are represented by $Y_{LD}$ \cite{R25};
\bea Y_{LD}\ga m_{b}, \hat s \dr&=& \frac{3}{\alpha^2}
\Bigg\{ - \frac{V_{cs}^* V_{cb}}{V_{ts}^* V_{tb}} \,C^{\ga 0 \dr}
- \frac{V_{us}^* V_{ub}}{V_{ts}^* V_{tb}}
\left[ 3 C_3 + C_4 + 3 C_5 + C_6 \right] \Bigg\} \nnb \\
&&\hspace{0.2in}\times \sum_{V_i = \psi \ga 1 s \dr, \cdots, \psi \ga
6 s \dr}
\ds{\frac{ \pi \kappa_{i} \Gamma \ga V_i \rar \ell^+ \ell^- \dr
M_{V_i} } {\ga M_{V_i}^2 - \hat s m_b^2 - i M_{V_i} \Gamma_{V_i} \dr
}}~, \eea
where $\kappa_i$ is a phenomenological parameter taken here to
be 2.3 so as to produce the correct branching ratio of $Br(B \to
J/\psi K^* \to K^* \ell \ell) = Br(B \to J/\psi K^*) Br(J/\psi \to
\ell \ell)$ \cite{R7}, and $C^{\ga 0 \dr} \equiv
3 C_1+C_2 + 3 C_3 + C_4 + 3 C_5 + C_6=0.362$. Contrary to $Y_{SD}$ the
long-distance contribution in $Y_{LD}$ has both weak and strong
phases. The weak phases follow from the CKM elements whereas the
strong phases come from the $\hat{s}$ values for which the $i$-th
charmonium states are on shell. Therefore, the Wilson coefficient
$C_{9}^{eff}(m_{b})$ has both weak and strong phases already in the
SM.

\par In this sense the Wilson coefficients $C_7^{eff}(m_b)$ and
$C_{10}(m_b)$ can not develop any strong phase, and thus, $\phi_7$ and
$\phi_{10}$ should necessarily originate from physics beyond the
SM. As such the phases of $\phi_7$ and $\phi_{10}$ can be chosen to
have a purely {\em weak} character.

%%%%%%%%%%%%%%%%%%%%%%%%%%%%%%%%%%%%%%%%%%%%%%%
%  Section 3

\section{Numerical analysis}

In this section we present our numerical results for the asymmetries
$A_{CP}$ for the $B \to K \mu^+ \mu^-$ decay. Note that the parameters
for the hadronic form-factors are taken from Table I. For values of
the Wilson coefficients in the SM we have used $C_3=0.011$,
$C_4=-0.026$, $C_5=0.007$, $C_6=-0.031$, $C_7^{eff}=-0.313$,
$C_9=4.344$, and $C_{10}=-4.664$. For further numerical analysis the
values of the new Wilson coefficients are needed, where we have varied
them in the range $-|C_{10}|<C_X<|C_{10}|$. The experimental value of
the branching ratio of the $B \to K(K^*) \ell^+ \ell^-$
decays\cite{R11,R12} and the bound on $Br(B \to \mu^+
\mu^-)$\cite{R26} suggest that this is the right order of
magnitude. It should be noted that the experimental 
results lead to strong restrictions on some of the Wilson
coefficients, namely $-2\leq C_{LL}$ and $C_{RL}\leq 2.3$, while the
remaining coefficients vary in the range $-|C_{10}|<C_X<|C_{10}|$. For
the remaining parameters we take $m_b=4.8$GeV, $m_c=1.35$GeV,
$m_B=5.28$GeV and $m_{K}=0.496$GeV.

\par For the kinematical interval the dilepton invariant mass is $4
m_{\ell}^{2} \leq q^2 \leq (m_B-m_{K})^2$ where the $J/\psi$ family of
resonances can be excited. The dominant contribution comes from the
three low-lying resonances $J/\psi, \psi^{'}, \psi^{''}$ in the
interval $8$GeV$^2\simlt q^2\simlt 14.5$GeV$^2$. In order to minimize
the hadronic uncertainties we will discard this subinterval in the
analysis below by dividing the $q^2$ region in to $low$ and $high$
dilepton mass intervals;
\bea
&&\mbox{Region I}:\ \ \ 4 m_{\ell}^{2} \leq q^2 \leq
8~\mathrm{GeV}^2,\nnb\\
&&\mbox{Region II}:\ \ \ 14.5~\mathrm{GeV}^2 \leq q^2 \leq
(m_B-m_{K})^2,
\eea
where the contribution of the higher resonances do still exist in the
second region.

%%%%%%%%%%%%%%%%%%%%%%%%%%%%%%%%%%%%%%%%%%
% Figure
\begin{figure}[t]
\epsfig{file=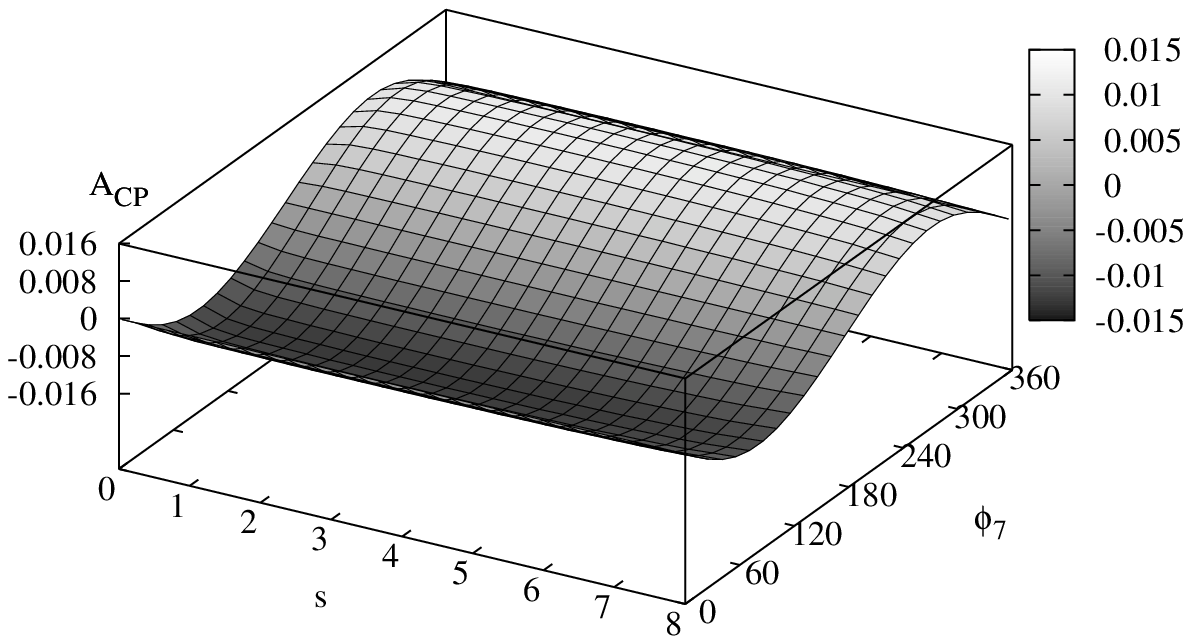,width=.48\textwidth}
\epsfig{file=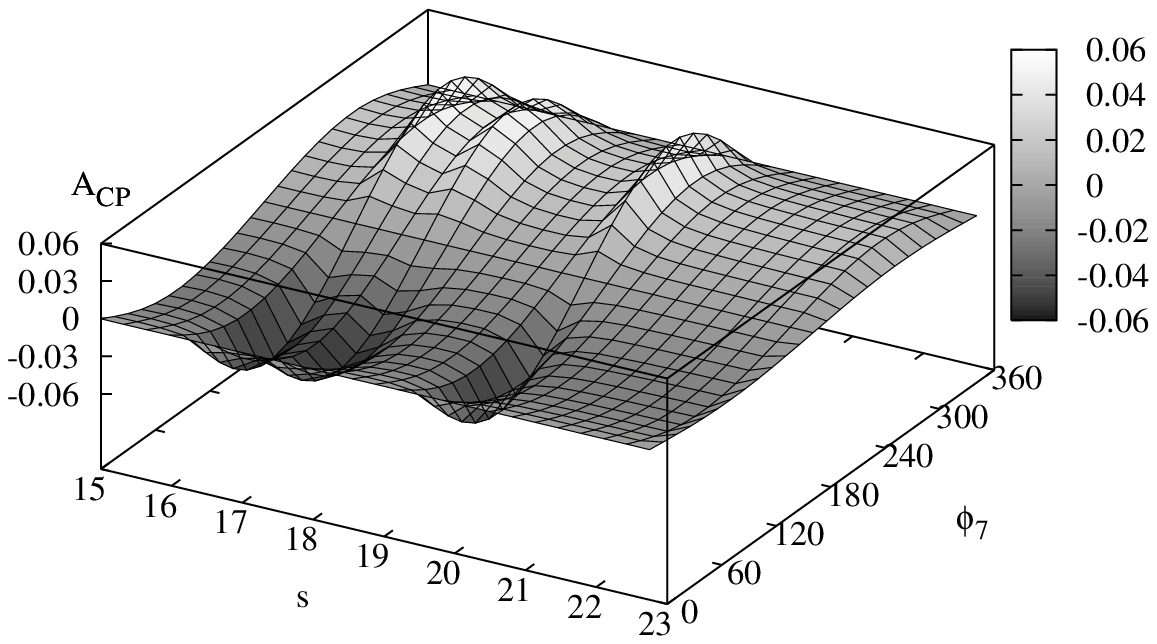,width=.48\textwidth}
\caption{\it Plot of the CP asymmetry in $B \to K \mu^+ \mu^-$ as a
function of the phase of $C_7 (\phi_7)$ and the dilepton invariant mass. The
left plot is for Region - I and the right plot is for Region - II. The other
Wilsons are taken to have their SM values.}
\label{fig:1}
\end{figure}
%%%%%%%%%%%%%%%%%%%%%%%%%%%%%%%%%%%%%%%%%%

\par As mentioned previously, we have analyzed the case where there
are four weak phases: $\phi_7$, $\phi_{10}$, $\phi_{RL}$ and
$\phi_{RR}$.

\par In fig. \ref{fig:1} we have presented $A_{CP}$ in the
$\phi_{7}$--$q^{2}$ plane for the $B \to K \mu^+ \mu^-$ decay for Region
I and Region II respectively. In Region I the CP
asymmetry is practically independent of $q^2$, becoming maximal in the
value for CP violation at $\phi_{7}=\pi/2$. In Region II, however, the
$q^2$ dependence is comparatively enhanced as the dominance of the dipole
coefficient is now reduced. Aside from this our figures suggest that
the CP asymmetry in Region II is four times larger than in Region I,
and this confirms our earlier expectation.

\par Since the CP asymmetry is dependent on $q^2$ and the new weak
phases there can appear some difficulties. The dependence of one of
the variables, for example $q^2$, can be removed by integrating over
$q^2$ in the allowed practical kinematical region, where the averaged
asymmetries could be measured more easily experimentally. Therefore we
shall now discuss only averaged CP asymmetries, which we define in the
following way. That is, our averaging procedure is defined by;
\beq 
\lla A_{CP} \rra = \frac{\displaystyle \int_{R_i} A_{CP} 
\frac{d \Gamma}{d q^{2}} d
q^{2}}{\displaystyle \int_{R_i} \frac{d \Gamma}{d q^{2}} d q^{2}}. 
\eeq
where $R_i$ means Region I or II.

%%%%%%%%%%%%%%%%%%%%%%%%%%%%%%%%%%%%%%
% Figure
\begin{figure}[hbt]
\begin{center}
\epsfig{file=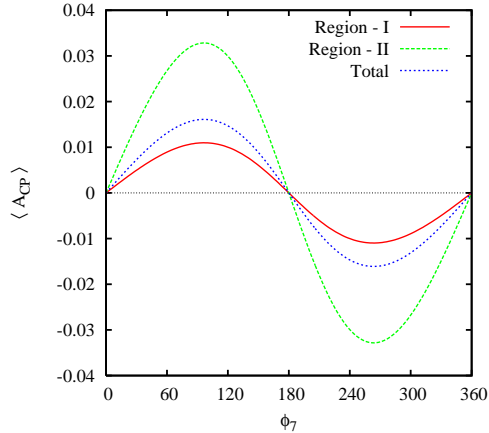,width=.5\textwidth}
\caption{\it The averaged CP asymmetry ($\langle A_{CP} \rangle$) in
the $B \to K \mu^+ \mu^-$ decay, where $C_7$ has a phase. The
remaining Wilsons are taken to have their SM values.} 
\label{fig:2}
\end{center}
\end{figure}
%%%%%%%%%%%%%%%%%%%%%%%%%%%%%%%%%%%%%%

%%%%%%%%%%%%%%%%%%%%%%%%%%%%%%%%%%%%%%
% Figure
\begin{figure}[hbt]
\begin{center}
\epsfig{file=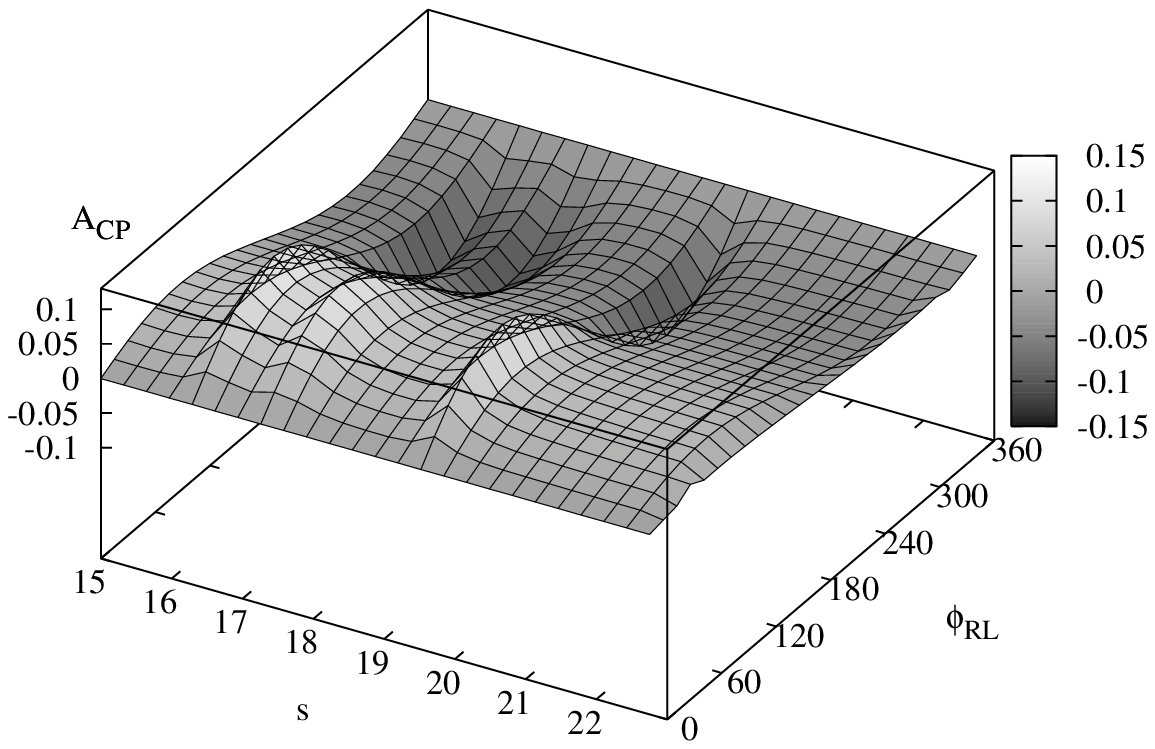,width=.5\textwidth}
\caption{\it A plot of the CP asymmetry in $B \to K \mu^+ \mu^-$ as a
function of the phase $\phi_{RL}$ and the dilepton invariant mass. In
this plot we have taken $|C_{RL}| = 2$ and with the other Wilsons as
having their SM values.} 
\end{center}
\label{fig:3a}
\end{figure}
%%%%%%%%%%%%%%%%%%%%%%%%%%%%%%%%%%%%%%

\par We now depict in fig. \ref{fig:2} the $\phi_7$ dependence of the 
averaged asymmetries $\langle A_{CP} \rangle$. From this figure it can
be observed that the average $\langle CP\rangle$ asymmetry can attain
values of $3\%$. Differences from zero of any value of $\langle A_{CP}
\rangle$ would be an unambiguous indication of the existence physics
beyond the SM.  

%%%%%%%%%%%%%%%%%%%%%%%%%%%%%%%%%%%%%%
% Figure
\begin{figure}[htb]
\begin{center}
\epsfig{file=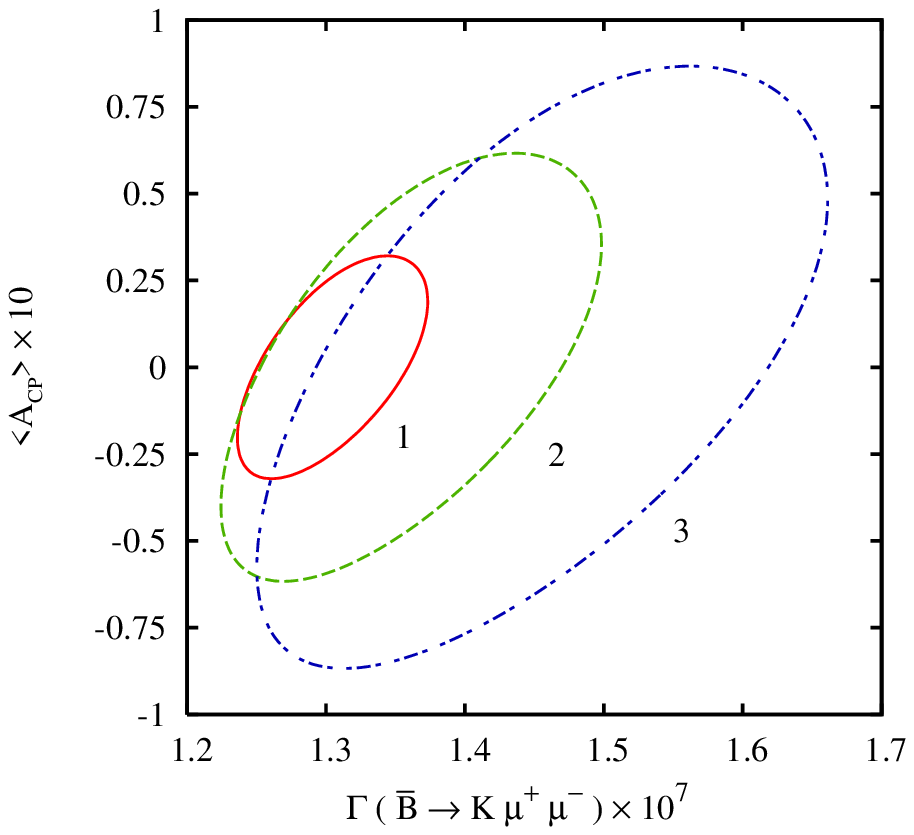,width=.5\textwidth}
\caption{\it The averaged CP asymmetry ($\langle A_{CP} \rangle$) in
the $B \to K \mu^+ \mu^-$ decay against the total branching ratio of
$B \to K \mu^+ \mu^-$. In this plot we have taken various values of
magnitudes of $C_{RL}$ (as stated in the figure) and varied the phase
in a range $0 \le \phi_{RL} \le 2 \pi$. We have plotted only Region -
II here.}  
\label{fig:3}
\end{center}
\end{figure}
%%%%%%%%%%%%%%%%%%%%%%%%%%%%%%%%%%%%%%

\par In fig. \ref{fig:3a} we have plotted the dependence of CP
asymmetry on the dilepton invariant mass and $\phi_{RL}$. In
fig. 5 we have shown the same kind of plot but for
$\phi_{RR}$. We have also shown the correlation of averaged CP
asymmetry and the integrated branching ratios. In fig. \ref{fig:3} the
variation of $\langle A_{CP} \rangle$ with integrated branching ratio
for $B \to K \mu^+ \mu^-$ for $C_{RL}$ is shown. In this figure we
have used three different values of $C_{RL}$ and have varied the phase
($\phi_{RL}$) in the range $0 \le \phi_{RL} \le 2 \pi$. All the
other Wilsons are taken to have their SM values. In a similar
graph, given in fig. \ref{fig:4}, we have varied $C_{RR}$.

%%%%%%%%%%%%%%%%%%%%%%%%%%%%%%%%%%%%%%
% Figure
\begin{figure}[htb]
\begin{center}
\epsfig{file=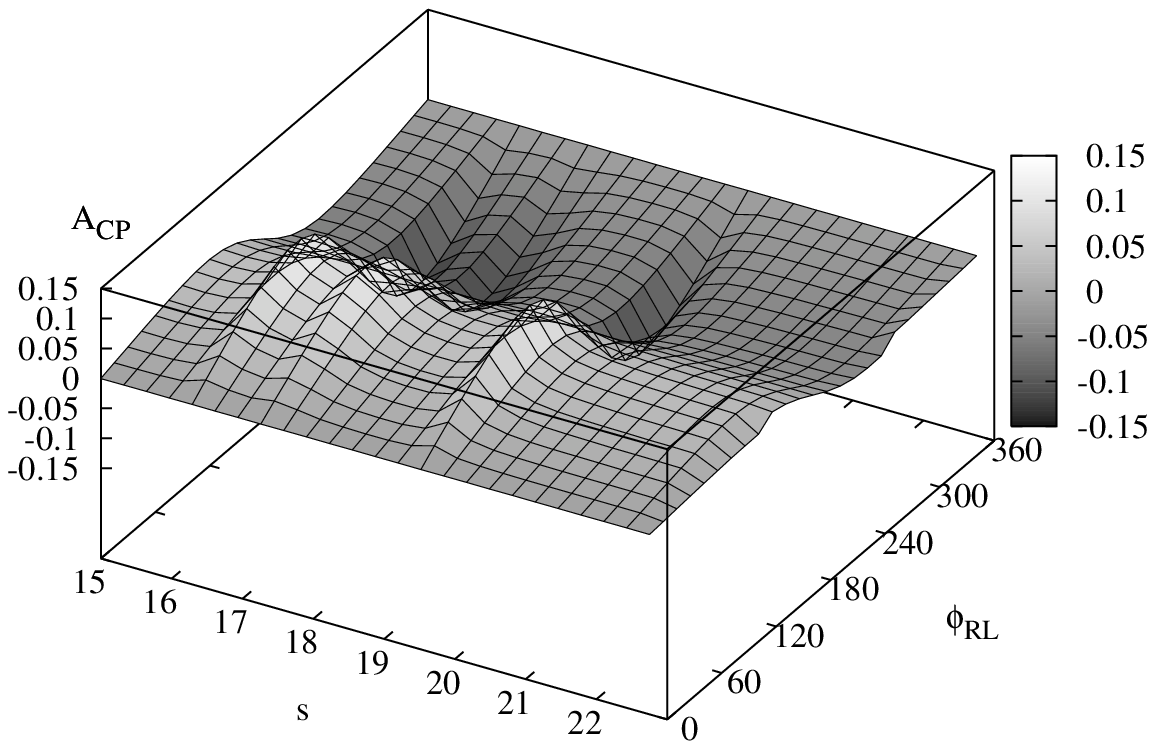,width=.5\textwidth}
\caption{\it A plot of the CP asymmetry in $B \to K \mu^+ \mu^-$ as a
function of the phase $\phi_{RR}$ and the dilepton invariant mass. In
this plot we have taken $|C_{RR}| = 2$ and with the other Wilsons as
having their SM values.} 
\end{center}
\label{fig:4a}
\end{figure}
%%%%%%%%%%%%%%%%%%%%%%%%%%%%%%%%%%%%%%

%%%%%%%%%%%%%%%%%%%%%%%%%%%%%%%%%%%%%%
% Figure
\begin{figure}[htb]
\begin{center}
\epsfig{file=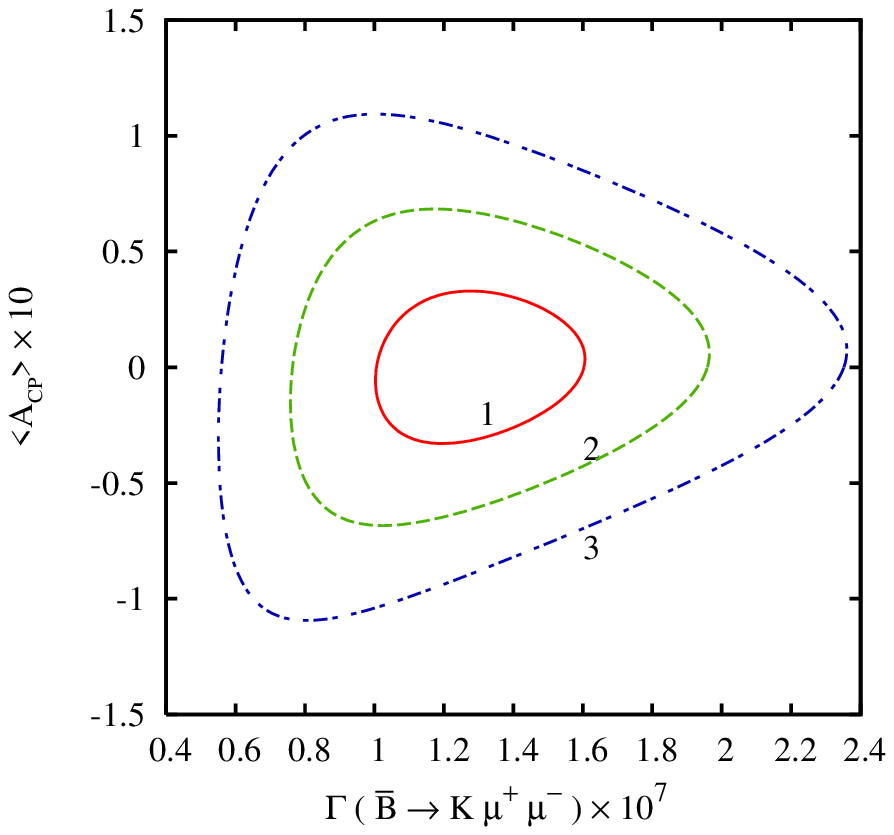,width=.5\textwidth}
\caption{\it The averaged CP asymmetry ($\langle A_{CP} \rangle$) in
the $B \to K \mu^+ \mu^-$ decay against the total branching ratio of
$B \to K \mu^+ \mu^-$. In this plot we have taken various values of
magnitudes of $C_{RR}$ (as stated in the figure) and varied the phase
in a range $0 \le \phi_{RR} \le 2 \pi$. We have plotted only Region -
II here.}    
\label{fig:4}
\end{center}
\end{figure}
%%%%%%%%%%%%%%%%%%%%%%%%%%%%%%%%%%%%%%

\par In the present work we have studied the sensitivity of the CP
violating asymmetry on the new weak phases appearing in the Wilson
coefficients. We have also observed that the CP asymmetry in Region II
is 4-5 times larger than that observed in Region I when we consider a
weak phase $\phi_7$. This can be understood in that in Region II
contributions coming from other operators become comparable with the
dipole operator $O_7$, where this operator is dominant in Region
I. Having obtained the averaged $\langle A_{CP} \rangle$ asymmetry we
obtained a maximal value of approximately 3\%. Note that an additional
weak phase in $C_{10}$ will not give rise to any CP asymmetry,
however, if non-standard\footnote{by non-standard we mean operators
which are not present within the SM} electroweak operators are
considered then the CP asymmetry in the region of high dilepton
invariant mass can reach a value of up to 10\%. 

\par As stated earlier, we can also, in principle, have weak phases in
scalar and pseudo-scalar operators. The presence of weak phases in
these operators can also substantially effect the CP asymmetry. The
popular extensions of the SM, such as SUSY and 2HDM, all predict the
existence of such operators. However, the magnitude of these Wilson
coefficients is predicted to be small when the lepton $\ell = e$ or
$\mu$. In the presence of these operators one also gets a non-zero
value for the FB asymmetry in $B \to K \ell^+ \ell^-$. The FB
asymmetry could provide another measure of CP
asymmetry\cite{Choudhury:1997xa} which has not been considered in this
work.   

\par The observation of CP asymmetry in $B \to K \ell^+ \ell^-$ would
not only tell us about the nature of weak phases but would also give
us an insight in to the structure of the effective
Hamiltonian. Therefore the measurement of the CP violating asymmetry
would provide us with an useful insight into the mechanism of CP
violation, which in turn would serve as a good test for physics beyond
the SM.   

%%%%%%%%%%%%%%%%%%%%%%%%%%%%%%%%%%%%%%%%%%%%%%%
%  Section : Acknowledgements

\section*{Acknowledgments}
The work of SRC and NG was supported by the Department of Science \&
Technology (DST), India under grant no SP/S2/K-20/99. The work of ASC
was supported by the Japan Society for the Promotion of Science
(JSPS), under fellowship no P04764.

%%%%%%%%%%%%%%%%%%%%%%%%%%%%%%%%%%%%%%%%%%%%%%%

\end{document}